\documentclass[twocolumn]{aastex62}

\usepackage{appendix}
\usepackage{amsmath,amssymb, amsthm,amstext}
\usepackage{natbib}
\usepackage{graphicx}
\usepackage{color}
\usepackage{array, enumerate, mathtools}
\usepackage{bm}
\usepackage{multirow}
\usepackage{braket}
\usepackage{txfonts}

\def\be{\begin{equation}}
\def\ee{\end{equation}}
\newcommand{\bL}{\bm L}
\newcommand{\ba}{\bm a}

\newcommand{\bJ}{\bm J}
\newcommand{\blambda}{\bm \lambda}
\newcommand{\risco}{r_{\rm isco}}

\submitjournal{ApJ}
\shorttitle{Probing the Growth of Massive Black Holes}
\shortauthors{Pan \& Yang}

\begin{document}
\title{Probing the Growth  of Massive Black Holes with Black Hole-Host Galaxy Spin  Correlations}

\author{Zhen Pan}
\affiliation{Perimeter Institute for Theoretical Physics, Waterloo, Ontario N2L 2Y5, Canada}
\email{zpan@perimeterinstitute.ca}

\author{Huan Yang}
\affiliation{Perimeter Institute for Theoretical Physics, Waterloo, Ontario N2L 2Y5, Canada}
\affiliation{University of Guelph, Guelph, Ontario N2L 3G1, Canada}
\email{hyang@perimeterinstitute.ca}

\begin{abstract}
Supermassive black holes (SMBHs) are commonly found at the centers of their host galaxies,
but their formation still remains an open question. In light of the tight correlation between the black hole (BH) mass
and the velocity dispersions of the bulge component of the host galaxy,
a  BH-host galaxy coevolution scenario has been established. Such description however still contains many theoretical uncertainties,
including the puzzels about the formation of BH seeds at high redshifts and the growth channel fueling
these seeds. In this work, we systematically analyze the signatures of different growth channels on massive BH (MBH) spins.
We show that different growth channels can be partially distinguished
with the magnitudes of MBH spins inferred from extreme-mass-ratio-inspirals
detected by the  Laser Interferometer Space Antenna. In addition, we propose to measure the correlation between the directions of
MBH spins and their host galaxy spins, which is possible for extreme mass-ratio inspirals happening in low-redshift galaxies ($z \le 0.3$).
With the inclusion of spin direction correlation different formation channels shall be significantly better constrained.
\end{abstract}
\keywords{Gravitational waves (678); Supermassive black holes (1663); Galaxy mergers(608)}

\section{Introduction}
Supermassive black holes (SMBHs) are commonly found at the centers of their host galaxies.
Empirical  correlations have also been extensively explored  \citep{Ferr2000,McConnell2013,Kormendy2013}
between the black hole (BH) masses, $M$, and different properties of their host galaxies,
including the velocity dispersion $\sigma_\star$ of bulge stars in the host galaxies.
The tight $M-\sigma_\star$ relation in combination with other correlations has inspired
an interpretation that BHs and their host galaxies coevolve by regulating each other's growth  \citep{Marconi2003,Kormendy2013}.
The coevolution scenario provides a framework that connects the galaxy evolution with BH activities.
However, there are still many important questions that  this coevolution scenario provides no definitive answers, especially the ones related to the SMBH formation, including the formation of BH seeds and the growth channels fueling these seeds.

BH seeds can be general classified as light seeds with masses in the range of $\sim (10^2, 10^3)M_\odot$
and heavy seeds within the range of $\sim (10^4,10^6)M_\odot$
\cite[see e.g.,][for reviews]{Rees1984,Latif2016, Haemmerl2020}. The light seeds are thought as  results of
collapse of metal-free population III stars \citep{Madau2001,Omukai2001,Abel2002,Heger2002}
and the heavy seeds are proposed to come from direct collapse of a massive protogalactic gas cloud
\citep{Begelman2006,Mayer2010,Di2012} or efficient merging stellar-mass compact objects in a
gas-rich environment \citep{Boco2020}.
Recently, \cite{Fragione2020} proposed that repeated mergers of stellar mass
BHs in nuclear star clusters can produce both light and heavy seeds depending on the cluster masses and densities.
Starting from the seeds at high redshifts,  BHs further accumulate masses by either merging with other BHs
or accreting gas, both of which seem to be compatible with available observations, including
the $M-\sigma_\star$ relation \citep{King2003, Volonteri2003, Marconi2004,Volonteri2005, King2006, Murray2005,Peng2007}.
As shown by \cite{Sesana2009} and \cite{Sesana2011}, both the seed formation and the BH growth history leave imprints on the mass
function of massive BH (MBH) binaries \footnote{We call BHs with masses $\sim(10^5, 10^7) M_\odot$ as MBHs.},
which can be probed from the LISA detection of
MBH coalescence from redshift $z= 10\sim 15$ to local universe \citep{Hughes2002,Barausse2015, LISA2017}.

Besides mass distributions, it is also natural to expect different models of seed formation and BH growth will also lead to different signatures on MBH spins, as discussed in \citep{Berti2008}.
MBH spins (both magnitudes and directions) can be accurately measured by LISA
from extreme-mass-ratio-inspiral (EMRI) events \citep{Huerta2009,Gair2017,Babak2017}, with the spin magnitudes
expected to be measured with fractional uncertainty $\sim(10^{-6},10^{-4})$ and the spin directions
expected to be constrained within $\sim (10^{-1}, 10^2)$ degree$^2$, which enables an accurate spin distribution reconstruction.
In this work, we propose to include another observable - the correlation between the spin directions of MBHs and their host galaxies, to further sharpen our ability to distinguish different formation models. Such observable relies on the host galaxy identification, which is possible for several percents of the EMRI events, and  galaxy spectroscopic surveys \citep{Bundy2015} to determine the galaxy spin orientation. \footnote{Galaxy spin orientations inferred from spectroscopic surveys have been used to probe the initial conditions in the early universe \citep{Motloch2020}. } We systematically analyze
the spin signatures of different growth channels assuming the natural light seeds scenario, and explore how likely it is to distinguish various channels with the spin information of MBHs using a Bayes method.\footnote{
MBHs stemming from light seeds accumulated almost all their masses via growth and their spins are completely determined by
their growth history. However, MBHs stemming from heavy seeds accumulated less masses from growth or did not grow at all,
therefore their spins depend more on the initial condition, i.e., the seed formation mechanism,
which is vaguely understood now but can be probed by high-redshift MBH mergers that are expected to
be detected by LISA \citep{LISA2017}, Einstein Telescope \citep{ET2020} and DECIGO \citep{DECIGO2020}.
The method of distinguishing different growth channels described in this paper equally applies to the heavy seed scenario,
as long as the seed formation mechanism is understood.}
In particular, for the first time we include
the MBH-host galaxy spin  correlations in the analysis, which turns
out to be a powerful probe to these growth channels.

The structure of the paper is organized as follows. In  Section \ref{sec:spins},
we outline the growth channels and model their
signatures on MBH spins and on the BH-host galaxy spin direction correlations.
In Section \ref{sec:LISA}, we introduce how well the MBH spins can be extracted
from the EMRI waveforms and how many host galaxies of EMRIs
can be identified in the LISA mission.
In Section \ref{sec:Bayes}, we show how to distinguish different growth models
from the spin information using the method of Bayesian model selection.
Some final remarks are given in Section \ref{sec:discuss}.

In this paper, we assume a flat $\Lambda$CDM cosmology with $\Omega_{\rm m} =0.3$ and $H_0=70$ km/s/Mpc, and we use geometrical units $G=c=1$.

\section{Growth Mechanisms of MBHs and Their Impacts on BH Spins}\label{sec:spins}

In this section, we review different growth channels of MBHs and obtain their corresponding implication on MBH spins, including the magnitude distribution of MBH spins and the MBH spin- galaxy spin correlation.

\subsection{Accretion}\label{sub:accretion}
\noindent{\it Coherent Accretion.} Accretion can be an efficient channel for MBHs to gain their masses \citep{Kawaguchi2003,Kawaguchi2004,King2006,Li2012}.
If a central BH is spun up with the large-scale gas fueling in a disk like configuration, the accretion is coherent.
In the standard thin-disk accretion, the BH could be spun up to a maximum value $a=0.998$ limited by the preferential accretion
of low-angular-momentum photons \citep{Thorne1974}.
In a magnetized disk, the equilibrium spin is $a \sim 0.95$ \citep{Gammie2004,Shapiro2005}.
In this work, we do not intend to distinguish the subtle differences arising from various assumptions on accretion physics,
and we choose to describe the spin magnitudes of MBHs with a coarse probability distribution $|a|\sim \mathcal N(1,0.05)$,
where $\mathcal N(\mu,\sigma)$ is a normal distribution with a mean value $\mu$ and a standard deviation $\sigma$
(here we use $\sigma=0.05$ to take account of typical variation of equilibrium BH spin magnitudes assuming different accretion
physics).
Following coherent accretion, the BH spin up to nearly extremal state, with its spin direction nearly aligned with the rotation direction of the accretion disk.
The initial spin magnitude and direction are no longer relevant after the BH mass increases by one or more e-folds.
As a result, the BH spin direction $\hat\ba$ should
be aligned with the rotation direction $\hat\bL$ of the large-scale gas disk.
In reality, the gas disk may be turbulent and sometimes clumpy with its local rotation direction
slightly off its mean value \citep[see e.g.,][]{Lima2017}. Though this variation is hard to calculate from first principle,
we may perform a rough estimate based on two quantities: the aspect ratio of the accretion disk $h:=H/r$ and the inclination
angle  $\iota_{\rm gas, star}$ between the gas disk and the stellar disk in the same galaxy. In the classical Mestel model of the gas disk \citep{Mestel1963}, $h$ is in the range of $\sim(0.05, 0.1)$. The inclination angles $\iota_{\rm gas, star}$ are
measured from galaxy spectroscopic surveys to be $\sim 10^\circ$  \citep{Krolewski2019}.
Therefore, we take $\delta\hat\bL\approx 10^\circ$ as a reference and
use an ansatz that $\cos^{-1}(\hat\ba\cdot\hat\bL) \sim \mathcal N(0, 10^\circ)$ in the following discussion.

\vspace{0.2cm}

\noindent{\it Chaotic Accretion.} The accretion is `chaotic' if it consists of many short chaotic episodes with different accretion direction in each one.
In this case, the distribution of BH spins is mainly determined by $\Delta M/M$, the fractional BH mass increase in each episode.
If $\Delta M/M\gtrsim 1$, The disk angular momentum in each episode is large enough to drive the BH to high spin no matter
what the initial spin is; while in episodes with $\Delta M/M\ll 1$ BH tends to spin down, because the BH mass increases linearly with the number of accretion episodes $N_{\rm acc}$ while the angular momentum gain is proportional $\sqrt{N_{\rm acc}}$ due to the random-walk cancellation \citep{King2006, King2008, Wang2009,Dotti2013,Volonteri2013,Liu2019,Zhang2019}.

In the beginning of each accretion episode, the BH spin is in general misaligned with the disk angular momentum,
and the inner part of the disk will be wrapped in a viscous timescale (known as the Bardeen-Petterson effect \citep{Bardeen1975}).
The wrapped disk will exert a torque onto the central BH and align or anti-align the BH spin with
 the angular momentum of the outer disk in a timescale
 \citep{Scheuer1996,Natarajan1998,King2005,Lodato2013, Gerosa2020}
 $t_{\rm align}\sim (M/\dot M)\alpha^{5/3} a^{2/3} h^{2/3}$
 assuming a standard $\alpha-$disk description \citep{Shakura1973},
 where $\dot M$ is the accretion rate and $h=H/r$ is the aspect ratio of the disk.
 After the alignment, the central BH will be spun up as accreting the gas.
Together with the accretion timescale $t_{\rm acc}=\Delta M/\dot M$, we obtain
\be
\frac{t_{\rm align}}{t_{\rm acc}} \simeq \frac{5\times 10^{-3}  |a|^{2/3}}{\Delta M/M} \left(\frac{\alpha}{0.1}\right)^{5/3}
\left(\frac{h}{0.1}\right)^{2/3}\ .
\ee
Therefore, we can safely ignore the short alignment period in calculating the final BH spin of each accretion episode
as long as $\Delta M/M \gg 5\times 10^{-3}  |a|^{2/3}$.

In each accretion episode,
the BH angular momentum $J$ changes with its mass $M$ as
\be\label{eq:Jevol}
dJ = l(a)M dM_{\rm gas} = \frac{l(a)}{e(a)} M dM\ ,
\ee
where $J=aM^2$, with $a$ being the dimensionless spin which is negative if the BH spin is anti-aligned with the angular momentum of the accretion disk; $l(a)$ and $e(a)$ are the specific angular momentum and specific energy of particles on the innermost stable circular orbit (ISCO) $\risco(a)$ with the following explicit forms \citep{Bardeen1972}
\be
\begin{aligned}
  e(a) &= \frac{\risco^{3/2}-2\risco^{1/2}+a}{\risco^{3/4}\left(\risco^{3/2}-3\risco^{1/2}+2a\right)}\ , \\
  l(a) &= \frac{\risco^{2}-2a\risco^{1/2}+a^2}{\risco^{3/4}\left(\risco^{3/2}-3\risco^{1/2}+2a\right)}\ ;
\end{aligned}
\ee
$dM_{\rm gas}$ is the mass element of accreted gas, a fraction $1-e(a)$ of which
is converted to radiation escaping to infinity and the remaining fraction $e(a)$ is absorbed by the BH.
As a result, we obtain the following simple evolution equation
\be\label{eq:1d}
da = \left[\frac{l(a)}{e(a)}-2a\right] d\ln M\ ,
\ee
where the specific angular momentum in retrograde accretion is larger than in direct accretion $l(a<0) > l(a>0)$,
leading to a larger spin magnitude change in retrograde accretion than in direct accretion
for a same mass increase $\Delta (\ln M)$.

\begin{figure}
\includegraphics[scale=0.6]{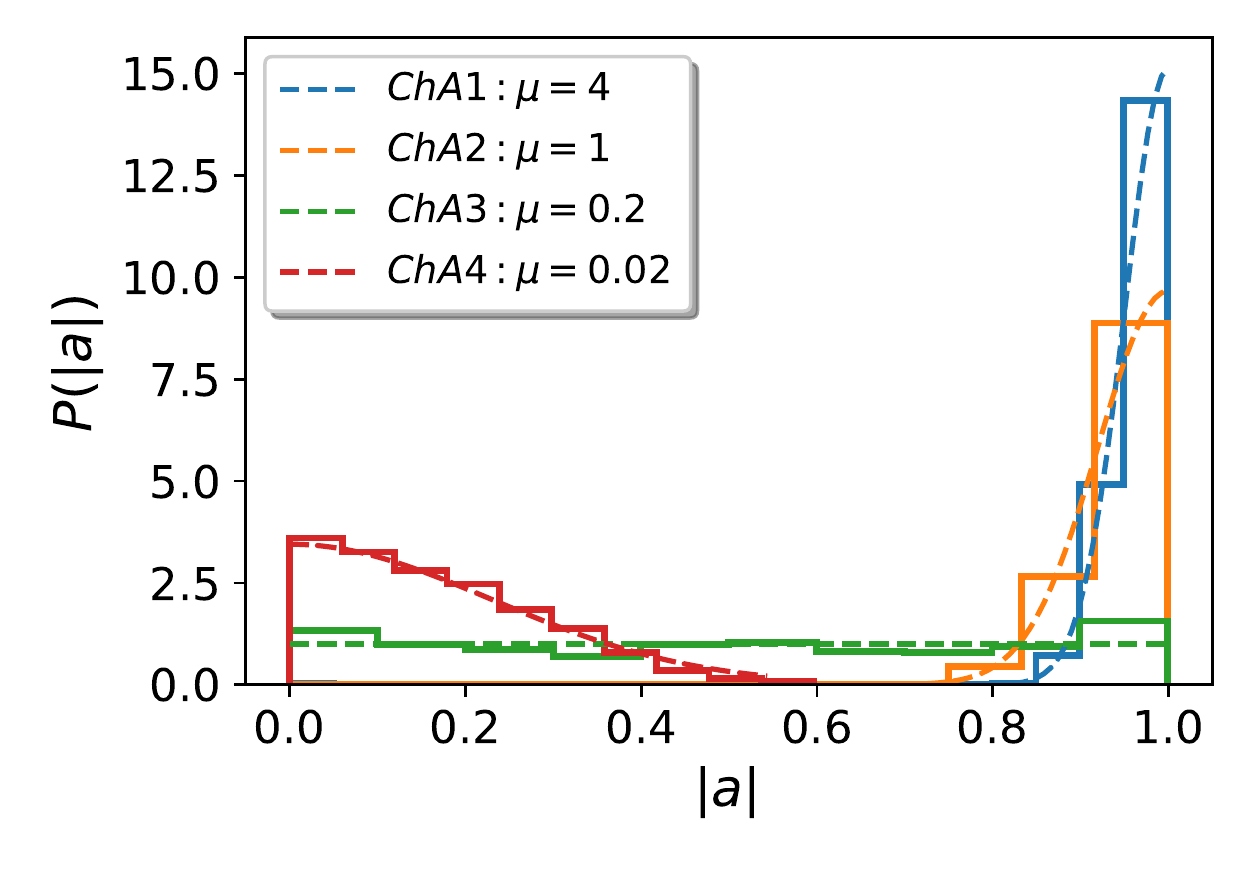}
\caption{\label{fig:chaotic} Distribution of BH spins driven by chaotic accretion with fractional mass increase $\Delta M/M\sim \mathcal N(\mu, \sigma)$ in each accretion episode, where solid lines are the simulation results and the dashed lines are the corresponding Gaussian fits. Here we take $\sigma=0.1\mu$ as an example, and larger $\sigma$, say, $\sigma=0.2
\mu$ only slightly broadens $P|_{\mu=1}$ and makes little change to remaining 3 distributions.}
\end{figure}

We consider a simple chaotic accretion model: in the beginning of each episode (after the short alignment process), the BH spin is assumed to be aligned or anti-aligned with the angular momentum of the accretion disk with equal chance and the BH increases by $\Delta M/M \sim \mathcal N(\mu, 0.1\mu)$.
To take account of the equilibrium BH spin as in the coherent accretion case, we enforce a spin distribution $P(a)\propto \mathcal N (1,0.05)$ for $a>0.9$.
In Fig.~\ref{fig:chaotic}, we show four example models of chaotic accretion (ChA1, ChA2, ChA3, ChA4 with $\mu = 4, 1, 0.2, 0.02$ respectively), where the spin distribution $P(|a|)$ is same to that of coherent accretion for $\mu\gtrsim 1$ because the
angular momentum gain is large enough to drive the BH to high spin whatever the initial spin is, and $P(|a|)$ peaks on zero for $\mu\ll 1$. In the case of $\mu\ll 1$, the BH will be spun up ($|a|\uparrow$) in direct accretion ($a>0$)
and will be spun down ($|a|\downarrow$) in retrograde accretion ($a<0$). As the spinning down is more efficient than the spinning up (see the explanation following Eq.~(\ref{eq:1d})), the net result is an equilibrium distribution $P(|a|)$ that peaks on $a=0$ and decreases with $|a|$.
In the case of chaotic accretion, we expect no BH-host galaxy spin direction correlation, i.e., $\hat\ba\cdot\hat\bL\sim \mathcal U(-1,1)$, where $\mathcal U$ is a uniform distribution.

\subsection{Mergers}\label{sec:merger}
\begin{figure}
\includegraphics[scale=0.5]{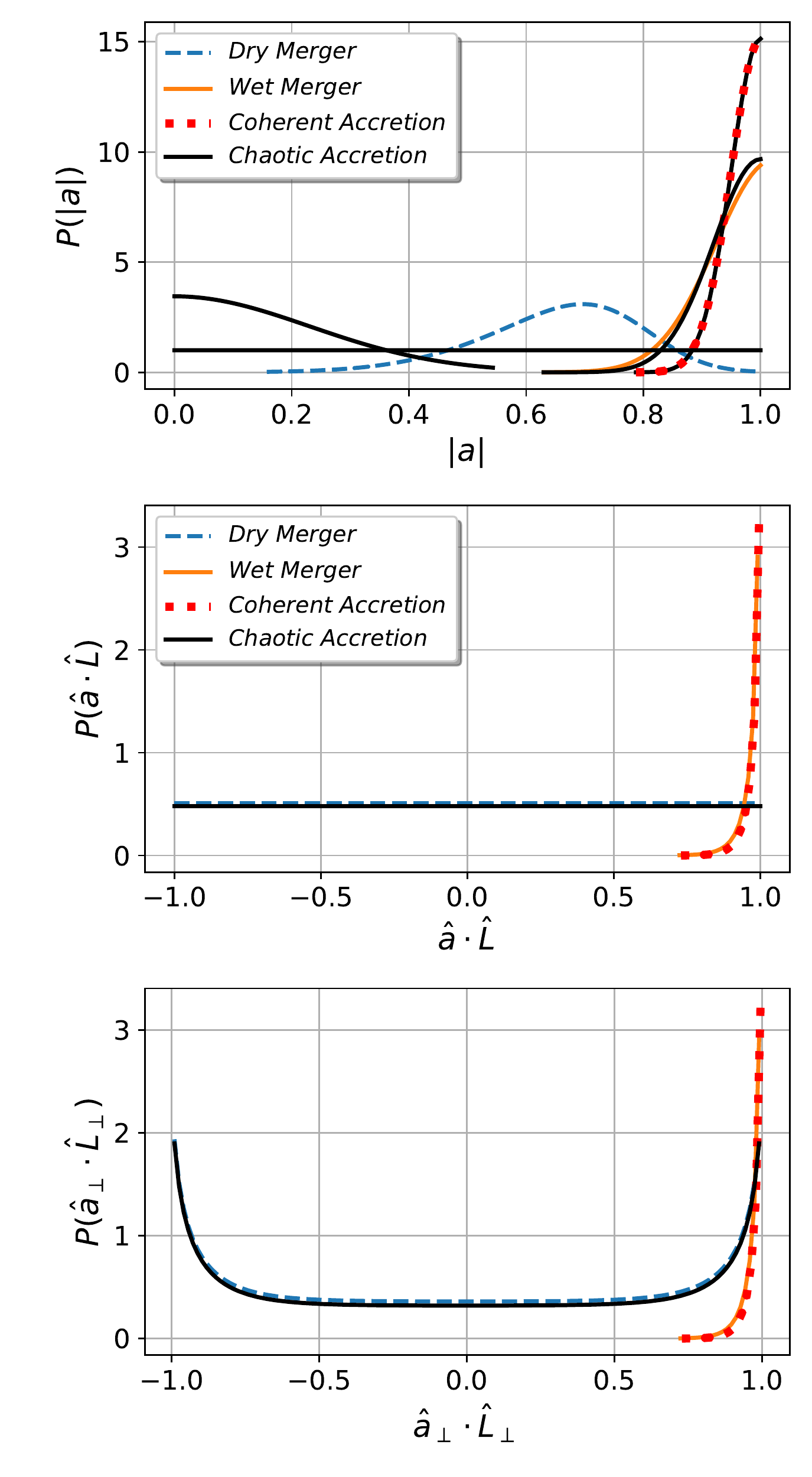}
\caption{\label{fig:spin} Upper panel: probability distribution functions of MBH spins of different growth channels,
where we show four chaotic accretion models with BH mass increase fraction $\Delta M/M\simeq 4,1, 0.2, 0.02$ in each accretion episode, respectively (also see Fig.~\ref{fig:chaotic}).
Middle panel: distributions of MBH-host galaxy spin direction correlations $\hat\ba\cdot\hat\bL$.
Lower panel: distributions of projected correlations $\hat\ba_\perp\cdot\hat\bL_\perp$,  where $\perp$
marks the projection to the plane perpendicular
to the LoS.}
\end{figure}

\noindent{\it Dry Mergers.}
MBH may merge following the merger of their host galaxies.  This is considered as a natural growth channel for MBHs considering
galaxies commonly harbor MBHs.  These mergers can be further classified as wet mergers (mergers in a gas-rich environment) and dry mergers  (mergers in a gas-poor environment) \citep[see e.g.,][for a review]{Colpi2014}. For dry mergers there are three main phases along the path to the final coalescence \citep{Begelman1980}:
(i) an early phase of pairing when MBHs migrate inwards driven by the dynamical friction with background stars,
until the two MBHs form a Keplerian binary \citep{Chandrasekhar1943, Begelman1980}; (ii) a  binary hardening phase when the binary separation decreases by ejecting stars of close encounters \citep{Yu2002,Milos2003}; (iii) a gravitational inspiral phase when the binary orbital decay is driven by the emission of GWs until the final coalescence.
For mergers of nearly equal mass BHs, the spins of remnant BHs peak around $a\simeq 0.69$, while mergers of small mass ratio BHs tend to produce remnant BHs with larger spin dispersion \citep{Barausse2009}. In the dry mergers of binary BHs, there is no apparent mechanism that aligns the BH spin directions  $\hat\ba_1,\hat\ba_2$ with their orbital direction $\hat \bL_{12}$
or aligns the orbital direction $\hat \bL_{12}$ with the host galaxy spin direction $\hat\bL$,
i.e., $\hat\ba_1,\hat\ba_2$ and $\bL_{12}$ are randomly oriented.
\citet{Berti2008}  investigated the cosmological spin evolution
of BHs driven by dry mergers (or isotropy mergers in their language) and they found the distribution of MBH spins peaks around $\sim 0.7$ with a long tail extending to small spins (we fit their histogram of BH spins with a skewed Gaussian distribution and plot in Fig.~\ref{fig:spin}).
The spin direction of the remnant BH should
also be randomly oriented with respect to the spin direction of the host galaxy, i.e., $\hat\ba\cdot\hat\bL\sim \mathcal U(-1,1)$.

\vspace{0.2cm}

\noindent{\it Wet Mergers.} If the two merging galaxies are gas rich, a stellar disk and a gas disk form in the remnant galaxy,
so that the two MBHs undergo roughly four different phases before the final coalescence \citep{Mayer2007}: (i)
an early pairing phase when MBHs wonder beyond the scope of the gas disk and migrate inwards
due to the dynamical friction with background stars; (ii) a  pairing phase when the motion of MBHs is influenced by the gravity of the gas disk (called circum-nuclear disk) and  migrate inwards driven by the torque from the density-wave excitations in the disk (similar to the Type I planet migration) \citep{Dotti2006, Dotti2007, Mayer2007,Colpi2009,Mayer2013}; (iii)
a hardening phase when the two MBHs form a Keplerian binary surrounded by a circum-binary disk
and migrate under the binary-disk coupling arising from two opposing actions that the binary tidal fields open
gaps in the disk whereas viscous torque fills the gaps (similar to the Type II planet migration) \citep{Armitage2002,Armitage2013}; (iv) and
an inspiral phase dominated by GW emission.

As a MBH with mass $M$ ram into the disk with an inclination angle $\iota$,
the inclination angle will be damped by the BH-disk interaction. Now we are to calculate the timescale of this process.
Assuming the disk volume density, surface density, circular velocity and sound speed at radius
$r$ are $\rho(r)$, $\Sigma(r)$, $V_{\rm d}(r)$ and $c_{\rm s}(r)$, respectively.  The disk aspect ratio $h$ is approximately
$h\simeq c_{\rm s}/V_{\rm d}$ and we take $h=0.1$ as a fiducial value. Gas bounded by the BH within radius $r_{\rm b} =GM/V_{\rm d}^2$ will be shocked and accelerated to roughly the same velocity of the BH. In a self-gravitating circum-nuclear disk, the BH speed is approximately $V_{\rm d}$.
Therefore the disk-BH interacting force is roughly
$F = \dot m_{\rm gas, acl} V_{\rm d} =\rho \pi r_{\rm b}^2  V_{\rm d}^2$
with $\dot m_{\rm gas, acl} $ being the amount of shocked and accelerated gas per unit time
and the inclination damping timescale is
\be
t_{\rm damp} = \frac{MV_{\rm d}\sin\iota}{F_\perp}\frac{\pi\sin\iota}{h}
=\frac{M}{\left(GM/V_{\rm d}^2\right)^2 \Sigma(r)} \frac{r\sin\iota}{V_{\rm d}}\ ,
\ee
where  $F_\perp = F\sin\iota$ is the component perpendicular to the disk and
factor $\pi\sin\iota/h$ takes account of the fact that only a fraction of the BH orbit
is inside the disk for $\iota > h$.
We consider a Mestel model of the circum-nuclear disk \citep{Mestel1963,Escala2005}.
The disk is self-gravitating and axisymmetric, with a constant rotational velocity $V_{\rm d}$,
which is related to the total disk mass $M_{\rm d}$ and the disk size $R_{\rm d}$ by $V_{\rm d}=\sqrt{GM_{\rm d}/R_{\rm d}}$.
The gas mass within radius $r$ is $M_{\rm gas}(r) = M_{\rm d}r/ R_{\rm d}$ and disk surface density at each radius is
$\Sigma(r) = M_{\rm d}/(2\pi R_{\rm d} r)$.
Using these relations, we find
\be
\begin{aligned}
  t_{\rm damp}
  &= \frac{M_{\rm gas}(r)}{M} \frac{2\pi r}{V_{\rm d}} \sin\iota \\
  & \stackrel{r=r_{\rm T} }{\sim} 5 \ {\rm kyr} \times \sin\iota \left(\frac{r_{\rm T}}{0.1 {\rm pc}}\right)
    \left(\frac{V_{\rm d}}{100\ {\rm km/s}}\right)^{-1} \ ,
\end{aligned}
\ee
where $r_{\rm T}$ is where $M_{\rm gas}(r)=M$, i.e., the transition radius from the circum-nuclear disk phase to
the circum-binary disk phase.
The timescale of inclination damping is much shorter than the typical migration timescale,
so we expect the gas disk and the binary BH orbit are coplanar at the end of the circum-nuclear disk phase,  $\hat\bL_{12}\cdot\hat\bL\approx1$.

In the circum-binary disk phase, the inner part of the disk will be warped if the BH spin $\hat\ba_{1,2}$ and the rotation direction of the disk $\hat \bL$ are misaligned. It is still not clear whether the interaction between the BH binary  and the warped disk can efficiently align them, though a number of studies have been performed previously
\citep[see e.g.,][]{Dotti2010,Maio2013,Lodato2013,Gerosa2020}. However, as shown by \citet{Barausse2009} and \citet{Hofmann2016}, the spin direction of the remnant BH $\hat\ba$ produced in the final coalescence is roughly aligned with the direction of the total angular momentum of the binary BH system at the beginning of the gravitational inspiral phase
to high precision with $\cos^{-1}(\hat\ba \cdot\hat\bJ_{\rm ini}) \sim \mathcal N(0, 5^\circ)$. Here $\bJ_{\rm ini}$ is the initial total angular momentum dominated by the orbital angular momentum $\bL_{12}$ at large seperations, and $\bL_{12}$  aligns with the spin direction of the gas disk $\bL$ as shown above. In combination with the intrinsic $\sim 10^\circ$ variation in the gas disk direction $\hat\bL$ (see section \ref{sub:accretion}), we obtain $\cos^{-1}(\hat\ba\cdot\hat\bL) \sim \mathcal N(0, \sqrt{(5^\circ)^2+(10^\circ)^2})$.
\citet{Berti2008} also simulated the spin magnitudes of MBHs from aligned mergers and we again fit the histogram they obtained with a skewed Gaussian distribution (Fig.~\ref{fig:spin}).

\vspace{0.2cm}
At this point, we have outlined the basic pictures of the different growth channels of MBHs.
The  corresponding BH spin distribution $P(|a|)$ and the BH-host galaxy spin direction
correlations $\hat\ba\cdot\hat\bL$ are also derived respectively.
However, $\hat \ba\cdot\hat\bL$ is not an ideal observable, because the 3D galaxy spin direction $\hat\bL$ is not
easy to measure, whereas its 2D projection $\hat\bL_\perp$ onto the plane perpendicular to the line of sight (LoS) can be accurately measured to $\approx 1^\circ$ precision in galaxy spectroscopic surveys (e.g., MaNGA, \citet{Bundy2015,Krolewski2019}).
Therefore, we also need to calculate the probability distributions of 2D spin direction correlation $\hat\ba_\perp\cdot\hat\bL_\perp$. For a given distribution $P(\hat \ba\cdot\hat\bL)$, we sample 32
data $\hat \ba\cdot\hat\bL$ points in concordance with the probability distribution; and for each pair of $\hat \ba, \hat\bL$,
we uniformly sample $1024$ directions of LoS, project the 3D directions $\hat\ba,\hat\bL$ onto the plane perpendicular to
the LoS and calculate the 2D correlation $\hat\ba_\perp\cdot\hat\bL_\perp$. We finally obtain a histogram of all $\hat\ba_\perp\cdot\hat\bL_\perp$ data points and fit it with a smooth distribution function
$P(\hat\ba_\perp\cdot\hat\bL_\perp)$ (Fig.~\ref{fig:spin}).
We find $P(\hat\ba_\perp\cdot\hat\bL_\perp)\approx P(\hat\ba\cdot\hat\bL)$ for Wet Merger and Coherent Accretion, and $P(\hat\ba_\perp\cdot\hat\bL_\perp)$ peaks on $\pm 1$ for Dry Merger and Chaotic Accretion with uniform $P(\hat\ba\cdot\hat\bL)$ due to projection distortion, i.e., there is a large chance of projecting $\hat\ba\cdot\hat\bL\approx 0$ to $\hat\ba_\perp\cdot\hat\bL_\perp\approx \pm 1$,
while the chance of projecting $\hat\ba\cdot\hat\bL\approx \pm1$ to $\hat\ba_\perp\cdot\hat\bL_\perp\approx 0$ is small.

\section{LISA Detection of EMRIs and  Host Galaxies Identification}\label{sec:LISA}
The expected EMRI rate depends on the mass function of MBH population at different redshifts,
the fraction of MBHs living in dense stellar cusps where stellar-mass BHs are produced, EMRI rate per MBH and properties of stellar-mass BHs in the cusps. \cite{Babak2017} quantified each of these astrophysical ingredients with semi-analytical models and calculated the
corresponding expected EMRI rates. They found tens to thousands of EMRIs per year should be detectable by LISA taking into account astrophysical uncertainties.
In particular, $\sim 6$ to $\sim 180$ low-redshift ($z\leq 0.5$) EMRIs are expected to be detected by LISA per year for the majority of the models considered \citep{Gair2017}.
For all detectable EMRIs, the typical fractional errors of intrinsic parameters, e.g., red-shifted masses and MBH spins, are
found in the range of $(10^{-6}, 10^{-4})$ \citep{Babak2017}. Luminosity distance can be constrained with precision $\sigma(\ln D_{\rm L})\approx \rho^{-1}$ and the median sky resolution is approximately $\sigma(\Omega_{\rm s})\approx 0.05(\rho/100)^{-5/2} \ {\rm degree}^2$, where $\rho$ is the signal to noise ratio (SNR) of the EMRI event \citep{McGee2020}.

\begin{figure*}
\includegraphics[scale=0.6]{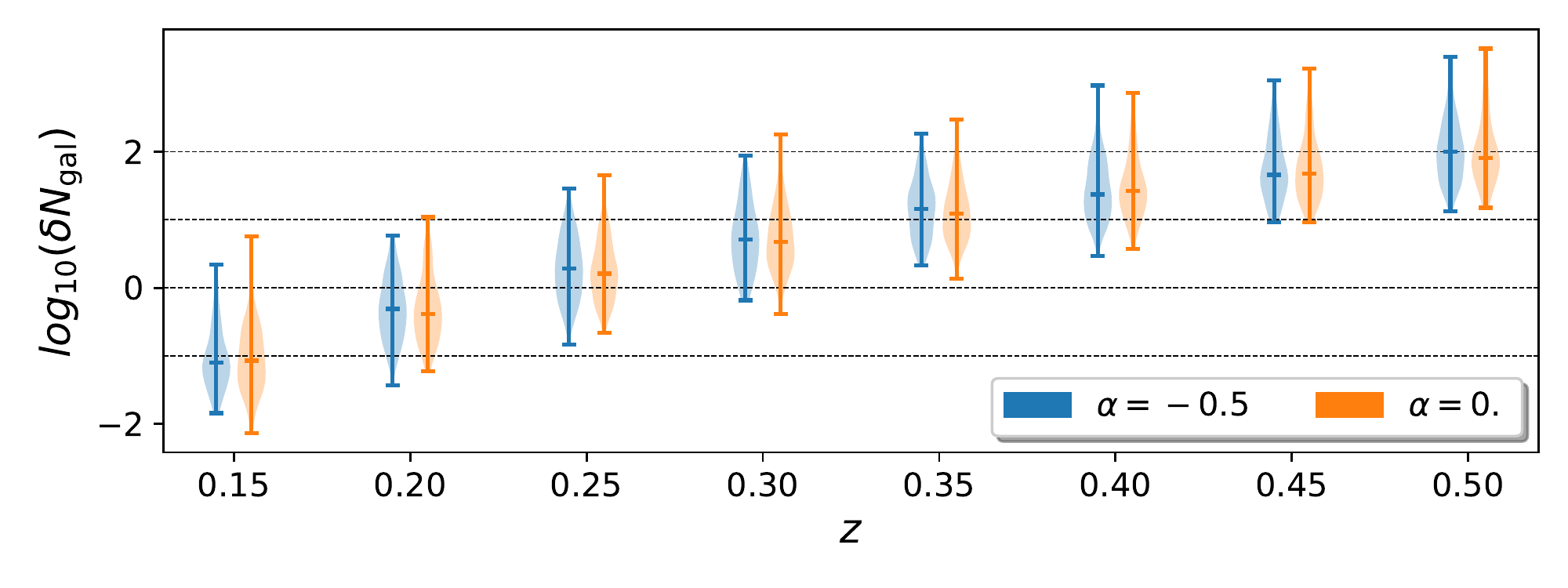}
\includegraphics[scale=0.6]{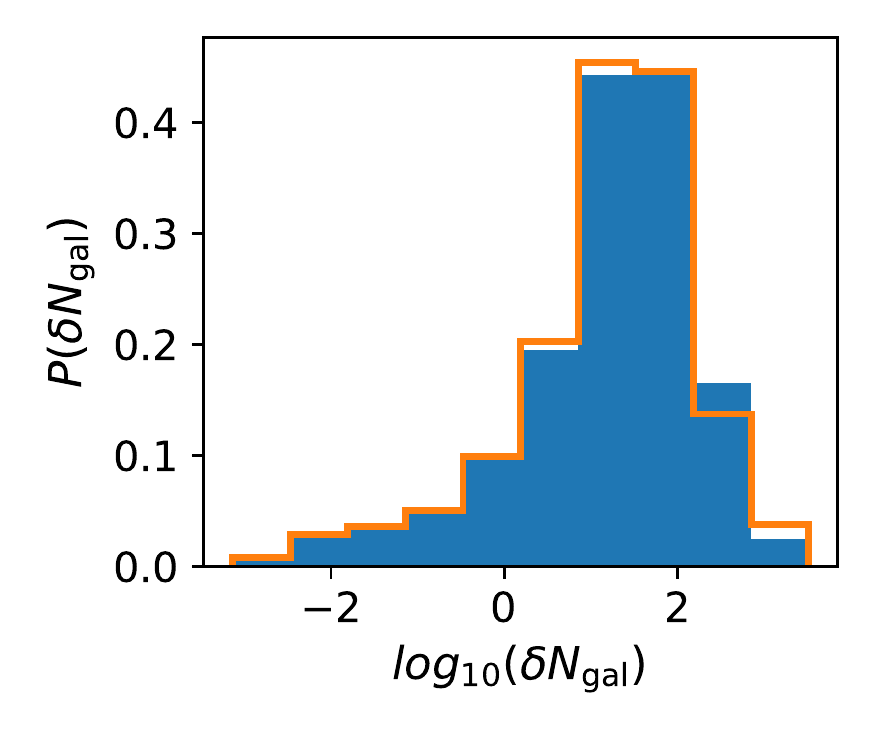}
\caption{\label{fig:pdf} Left panel shows the distribution of expected number of galaxies $\delta N_{\rm gal}$
in each error box of the LISA detected EMRIs. Right panel shows the histogram of $\log_{10}(\delta N_{\rm gal})$ for all detected EMRIs with redshift $z<0.5$, where $P(\delta N_{\rm gal} \leq 1) = 13\%$,
$P(1<\delta N_{\rm gal} \leq 10) = 21\%$ and $P(10<\delta N_{\rm gal} \leq 100) = 48\%$  for mass function with power index with $\alpha = -0.5$,
and the histogram for mass function with $\alpha = 0$ is approximately the same. }
\end{figure*}

Based on these results, we expect there is a fraction of low-redshift EMRIs that can be localized to small 3D error boxes  containing a single galaxy. Therefore the host galaxies of these EMRIs can be identified from the corresponding  LISA observation.
Following the approach in  \citep{Babak2017},  we consider a power-law mass function of the EMRI population with a redshift-independent EMRI rate $R_0$ at redshift
$z\leq 0.5$,
\be\label{eq:massfunc}
\frac{dR_0}{d\ln M}\propto M^\alpha \ ({\rm for}\ 10^5 M_\odot<M<10^6 M_\odot)
\ee
with power index $\alpha = -0.5$ or $0$. At each redshift $z$, we sample $128$ EMRIs with the MBH mass sampled from Eq.~(\ref{eq:massfunc}), the MBH spin chosen as $a=0.98$, the companion BH mass set to be $m=10 M_\odot$, the binary orbital eccentricity
at plunge being $e_{\rm p}= 0.1$, the luminosity distance $D_{\rm L}(z)$, and 8 randomly sampled angles (including the source sky localization angles $(\theta_s, \phi_s)$
and the MBH spin direction angles $(\theta_a, \phi_a)$) that uniquely determine the binary configuration at coalescence
(see \citet{Chua2015,Chua2017} for details). For each EMRI, we model its GWs
with the Augment Analytic Kludge (AAK) waveform (\cite{Chua2017}) and record the
time-domain waveforms $h_{\rm I}(t)$  and $h_{\rm II}(t)$ in the last two years
 before coalescence (${\rm I}$ and ${\rm II}$ mark the two orthogonal LISA channels), where \citep{Barack2004,Rubbo2004}
\be
\begin{aligned}
  h_{\rm I}(t) &= h_+(t)F_{\rm I}^+(t)+ h_\times(t)F_{\rm I}^\times(t), \\
  h_{\rm II}(t) &= h_+(t)F_{\rm II}^+(t)+ h_\times(t)F_{\rm II}^\times(t),
\end{aligned}
\ee
with $h_{+,\times}$ being the waveforms of two polarizations and $F^{+,\times}_{\rm I,II}$ are the corresponding
detector antenna patterns of the two channels.

The SNR of EMRI GWs is calculated as
$\rho =\sqrt{\braket{h_{\rm I}|h_{\rm I}}+\braket{h_{\rm II}|h_{\rm II}}}$ with the inner product defined as
\be
\braket{u|v} := 4 \int_0^\infty \frac{\mathcal R[u(f)v^*(f)] }{P_{\rm n}(f)} df\ ,
\ee
where $\mathcal R$ denotes the real part and $P_{\rm n}(f)$ is the combination of  one-side spectral density of
the LISA detector noise and the residual foreground of unresolvable
galactic binary white dwarfs \citep{LISA2017,Robson2019}.
To keep consistent with the criteria used in \citet{Babak2017}, we choose $\rho=20$ as the threshold of EMRI detections. For EMRIs with $\rho>20$, we forecast
the model parameter constraints with Fisher matrix
\be
F_{ij} = \braket{\frac{\partial h_{\rm I}}{\partial \lambda_i}|\frac{\partial h_{\rm I}}{\partial \lambda_j}}
+\braket{\frac{\partial h_{\rm II}}{\partial \lambda_i}|\frac{\partial h_{\rm II}}{\partial \lambda_j}}\ ,
\ee
where $\lambda_{i,j}\ (i,j=1,...,13)$ are the EMRI model parameters briefed in the previous paragraph.
The 1-$\sigma$ uncertainty of parameter $\lambda_i$ is $\sigma(\lambda_i) = \sqrt{C_{ii}}$ with $C_{ij} := F^{-1}_{ij}$
being the covariance matrix.

The volume of the 3D error box is
$\delta V = r^3(z)\times \pi \sin\theta_s \sqrt{C_{\theta_s\theta_s}C_{\phi_s\phi_s}-C_{\theta_s\phi_s}C_{\theta_s\phi_s}} \sigma(\ln D_{\rm L}) $, where $r(z)=\int_0^z\frac{dz}{H(z)}$ is the comoving distance from the EMRI source to the earth, with $H(z)$ being the Hubble expansion rate. Inside the error box, we expect to see $\delta N_{\rm gal}$ galaxies, with  $\delta N_{\rm gal} = \bar n_{\rm gal}\delta V$.
Here the average
number of galaxies per comoving volume is chosen as  $ \bar n_{\rm gal}=10^{-2}$ Mpc$^{-3}$  \citep[consistent with][]{Kuns2019}. In Fig.~\ref{fig:pdf}, we show the $\delta N_{\rm gal}$ distribution
at each redshift, where we find $13\%$ of the detectable EMRIs with $z<0.5$  can be traced back to their host galaxies,
i.e., $\sim 8$ to $\sim 234$ EMRIs and their host galaxies can be identified by LISA in the maximum
mission duration $\approx 10$ years \citep{LISA2017}.

\begin{figure*}
\includegraphics[scale=0.7]{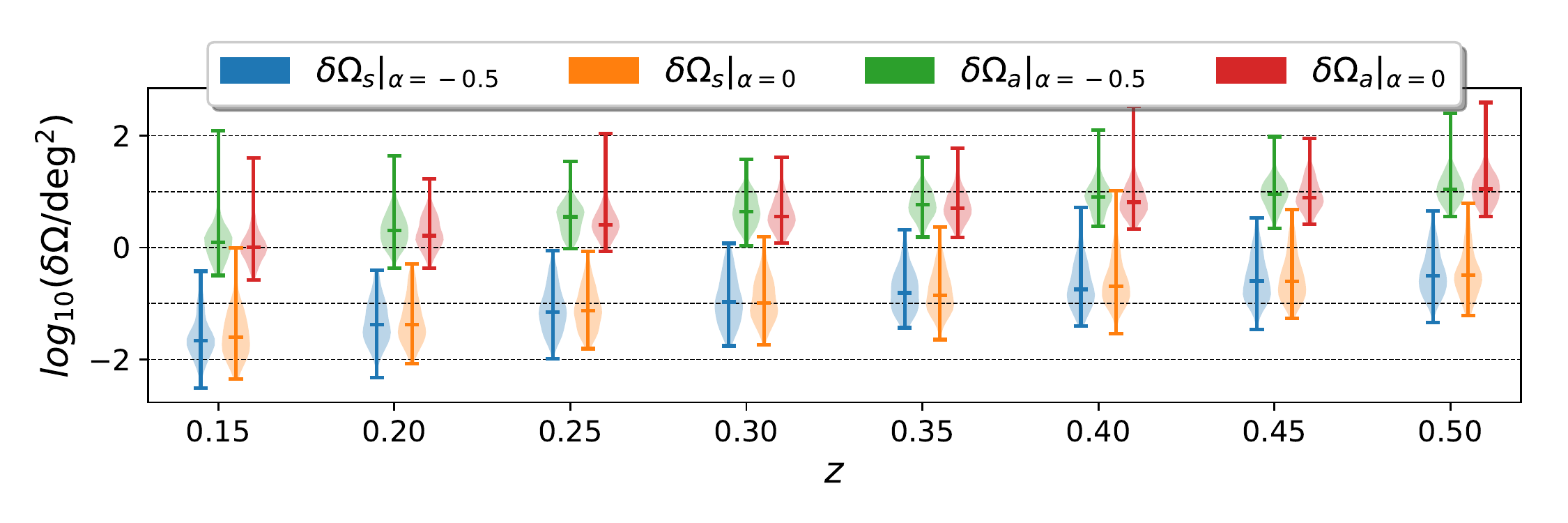}
\caption{\label{fig:Omega} Distributions of sky resolution $\delta\Omega_{\rm s}$ and spin direction $\hat\ba$ resolution  $\delta\Omega_a$  of detectable EMRIs at each redshift $z$.}
\end{figure*}

In Fig.~\ref{fig:Omega}, we also show the sky resolution $\delta\Omega_{\rm s}$ and spin direction $\hat\ba$ resolution  $\delta\Omega_a$  of detectable EMRIs at each redshift,  where $\delta\Omega_{\rm s}:=2\pi
\sin\theta_s \sqrt{C_{\theta_s\theta_s}C_{\phi_s\phi_s}-C_{\theta_s\phi_s}C_{\theta_s\phi_s}}$
and  $\delta\Omega_a$ is defined in a similar way. They will be used in estimating the data errorbars of MBH spins and
MBH-host galaxy spin direction correlations.

\section{Bayesian Model Selection}\label{sec:Bayes}
As explained in the previous two sections, different growth channels will leave different imprints on
MBH spins and on MBH-host galaxy spin direction correlations. The former can be measured
by LISA from EMRIs, and the latter can  be measured by LISA in combination with galaxy spectroscopic surveys.
In this section, we will quantitatively explore how likely these channels can be distinguished given data of MBH spins and MBH-host galaxy spin correlations.

According to Bayes theorem, the posterior probability $\mathcal P({\blambda}|D, \bm m)$
for the parameters $\blambda$ of a model $m$ given data $D$ is related to the likelihood $\mathcal P(D|\blambda, \bm m)$
of seeing data $D$ under model $\bm m$ with model parameter $\blambda$  by
\be
\mathcal P({\blambda}|D, \bm m) = \frac{\mathcal P(D|\blambda, \bm m)\mathcal P(\blambda|\bm m)}{\mathcal E(D|\bm m)}\ ,
\ee
where  $\mathcal P(\blambda|\bm m)$ is  the prior and $\mathcal E(D|\bm m) = \int \mathcal P(D|\blambda, \bm m)\mathcal P(\blambda|\bm m) d\blambda$ is
the evidence of model $\bm m$ given data $D$. To determine  the (de)preference of  models $m_1$ over $m_2$ based on data $D$, we calculate the Bayes factor
\be\label{eq:Bayes}
\mathcal B^{m1}_{m_2}(D) = \frac{\mathcal E(D|m_1)}{\mathcal E(D|m_2)}\ .
\ee
If $\mathcal B^{m1}_{m_2} > 1$, $m_1$ is a better model than $m_2$; and if $\mathcal B^{m1}_{m_2}< 1$, $m_2$ is better.
According to Jeffreys' evidence scale, $\ln\mathcal B^{m1}_{m_2} >5$ is interpreted as $m_1$ is overwhelmingly better than $m_2$ and equivalently
$\ln\mathcal B^{m1}_{m_2} < -5$ as $m_2$ is overwhelmingly better.

In the context of this work, there is no free parameter in the considered models (Fig.~\ref{fig:spin}),
thus the evidence calculation is simplified as $\mathcal E(D|\bm m) = \mathcal P(D|\bm m)$.
In order to calculate the likelihood $\mathcal P(D|\bm m)$,
we first divide data into $N_{\rm B}$ bins ${B_i}$ $(i=1,...,N_{\rm B})$,
and count the number of events in each bin $\{n_i\}$. Each model will predict an average probability of events occuring
in each bin $P_i(\bm m):=\int_{B_i} P(d|\bm m)\ {\rm d} d$ (see Fig.~\ref{fig:spin} for the probability distribution functions $P(d|\bm m)$ ). Events in different bins are independent and the number of events in each bin should satisfy Possion distribution with an average probability $P_i$, therefore the likelihood is written as \citep{Gair2010,Gair2011}
\be
\begin{aligned}
  \mathcal P(D|\bm m)
  &= \prod_{i=1}^{N_{\rm B}}\frac{\left[N_{\rm D}(\bm m)P_i(\bm m)\right]^{n_i}e^{-{N_{\rm D}(\bm m)P_i(\bm m)}}}{n_i!}\ ,\\
  &= \left[N_{\rm D}(\bm m)\right]^{\sum n_i}e^{-N_{\rm D}(\bm m)}\prod_{i=1}^{N_{\rm B}}\frac{\left[P_i(\bm m)\right]^{n_i}}{n_i!}\ ,
\end{aligned}
\ee
where $N_{\rm D}(\bm m)$ is the number of events predicted by model $\bm m$, and $N_{\rm D}(\bm m)P_i(\bm m)$ is
the expected number of events in bin $B_i$, and we have used the normalization condition $\sum P_i(\bm m) = 1$ in the second equal sign. That is to say, different models predict not only different distributions $P_i(\bm m)$
of data but also different occurrences $N_{\rm D}(\bm m)$ of data, both of which contribute to model selections.
In fact, the total number $N_{\rm D}(\bm m)$ is commonly more uncertain than the distribution $P_i(\bm m)$.
In this paper, we will only use the distribution  $P_i(\bm m)$ information to distinguish different channels, i.e.,
we set $N_{\rm D}({\bm m})$ to be same for different models, and we obtain
\be\label{eq:likeli}
\mathcal P(D|\bm m)
= {\rm const}\times \prod_{i=1}^{N_{\rm B}}\frac{\left[P_i(\bm m)\right]^{n_i}}{n_i!}\ .
\ee
In reality, any data point is subject to some measurement uncertainty,
$D=\{d^j\pm \delta d^j\}$. We model the true value of event $j$ with
a probability distribution $\mathcal N(d-d^j, \delta d^j)$ and assign a fractional occurence
$\int_{B_i} \mathcal N(d-d^j, \delta d^j) \ {\rm d} d $ into each bin $B_i$.
In the continuum limit (small bins limit), Eq.~(\ref{eq:likeli}) simplifies as
\be\label{eq:likeli_conti}
\mathcal P(D|\bm m)
= {\rm const}\times \prod_{j=1}^{N_{\rm D}} P_j(\bm m)\ ,
\ee
where $P_j(\bm m)= \int P(d|\bm m) \mathcal N(d-d^j, \delta d^j)  \ {\rm d} d $.

In our case, data $D$ includes both the MBH spins $D_1=\{|a^j|\pm \delta_a^j\}\ (j=1,...,N_1)$ of the EMRIs detected by LISA
and the MBH-host galaxy spin correlations $D_2=\{\hat\ba_\perp^j\cdot\hat\bL_\perp^j\pm \delta_{aL}^j\}\ (j=1,...,N_2)$,
where the error bar $\delta_a^j$ is obtained from Fisher forecasts explained in the previous section, while
$\delta_{aL}^j$ depends on the spin direction uncertainty $\delta\Omega_a$,
the sky location uncertainty $\delta\Omega_s$,
the angle between the spin direction and the LoS $\theta_{a, \rm LoS}$, and the uncertainty $\delta_L^j$ of $\hat{\bm L}_\perp^j$. As shown in \citet{Bundy2015} and \cite{Krolewski2019}, the galaxy spin direction $\hat\bL_\perp^j$ can be measured
with uncertainty $\delta_L^j\approx 1^\circ$. From Fig.~\ref{fig:Omega}, the sky resolution $\delta\Omega_s$ of LISA turns out be $\sim 30$ times better than than of MBH spin direction $\delta\Omega_a$. As a result, we find $\delta_{aL}^j$ is dominated by the uncertainty of MBH spin direction $\delta\Omega_a^j$.
In terms of azimuthal angles, $\hat\ba_\perp\cdot\hat\bL_\perp \pm \delta_{aL} = \cos^{-1}(\phi_{aL}\pm\delta\phi_{aL})$,
we have $\delta\phi_{aL}\approx \sqrt{\frac{\delta\Omega_a}{2\pi}}/\sin^2\theta_{a,\rm LoS}$, where $\phi_{aL}$ is the angle
between $\hat\ba_\perp$ and $\hat\bL_\perp$, and $\delta\phi_{aL}$ is its uncertainty.

\begin{table}
\begin{center}
  \begin{tabular}{ l |  c  c | c c }
    $m_1$\textbackslash $m_2$ &   \multicolumn{2}{c|}{WM} &  \multicolumn{2}{c}{CoA} \\
    \hline
    DM    & $-195\pm16$ & ($-210\pm16$)& $-267\pm12$ & ($-283\pm12$) \\
    WM    & & & $-10\pm3.8$ & ($-10\pm3.8$) \\
    CoA   &  $-35\pm14$ &($-35\pm14$)&  \\
    ChA1  & $-35\pm14$ & ($-50\pm14$) & $0\pm0$ &($-15\pm2.2$) \\
    ChA2  & $-1.2\pm1.6$ & ($-16\pm2.7$) & $-6.8\pm3.6$ & ($-22\pm3.8$)\\
    \hline
    \hline
  \end{tabular}
  \caption{\label{tab} Bayes factors $\ln\mathcal B^{m_1}_{m_2}$ of model $m_1$ relative to $m_2$, where
  ChA1= ChA$|_{\Delta M/M\simeq4}$ and ChA2= ChA$|_{\Delta M/M\simeq1}$. In each column,
  we take $m_2$ as the true underlying model and generate data $D_1=\{|a^j|\}\ (j=1,...,60)$ and data
  $D_2=\{\ba^k_\perp\cdot\bL^k_\perp\}\ (k=1,...,8)$.
  Numbers outside parentheses are the  Bayes factors $\ln\mathcal B^{m_1}_{m_2}(D_1)$
  and numbers inside are  $\ln\mathcal B^{m_1}_{m_2}(D_1+D_2)$.
  We do not include models ChA3 and ChA4
  simply because they predict distinct spin distributions that can easily be indentified by eye.}
\end{center}
\end{table}

To illustrate how well different growth channels can be distinguished from each other based on the MBH spin magnitudes data $D_1$
and MBH-host galaxy spin correlation data $D_2$, we take a model $m_2$ as the true underlying model and generate 256 realizations of $D_1$ and $D_2$ sampled from the corresponding distributions shown in Fig.~\ref{fig:spin} and \ref{fig:pdf}.
We conservatively take the data sizes of $N_1=60$ and $N_2=8$. For each realization of data, we can calculate the Bayes factors $\mathcal B^{m_1}_{m_2}$ of model $m_1$ relative to $m_2$ given data $D_1$ or given both data $D_{1,2}$ using Eqs.~(\ref{eq:Bayes},\ref{eq:likeli_conti}).
In Table \ref{tab}, we list the results of $\ln \mathcal B^{m_1}_{m_2}$, with $m_2$ being Wet Merger or Coherent Accretion, and $m_1$ being Dry Merger (DM), Wet Merger (WM), Coherent Accretion (CoA) or Chaotic Accretion (ChA).
If WM is the true underlying model, we find it can be distinguished from DM/CoA/ChA1 with overwhelming evidence
($|\ln \mathcal B^{m_1}_{m_2}|>5$) at $12/2.1/2.1\ \sigma$ confidence level, while is indistinguishable from ChA2 with data $D_1$ only.
Adding data $D_2$ into consideration, the Bayes factor contrasts $|\ln \mathcal B^{m_1}_{m_2}|$ increase by $15$ for $m_1=$ DM/ChA1/ChA2,
and WM can be distinguished from DM/CoA/ChA1/ChA2 with overwhelming evidence at $13/2.1/3.2/4.1 \ \sigma$ confidence level. Similar behaviors are found if the underlying model is CoA.

\section{Discussion}\label{sec:discuss}
Both the formation of BH seeds and the growth history of MBHs leave imprints on the mass function of MBHs,
on the distribution of MBH spins and on the MBH-host galaxy spin direction correlations.
The mass function can be reconstructed from LISA detected MBH coalescence from high redshift to local universe,
the spin distribution can be measured from LISA detected EMRIs and the spin direction correlations can be measured
in combination with galaxy spectroscopic surveys. In this paper, we show that different growth channels are partially
 distinguished from the MBH spins, and can be significantly better distinguished in combination with even a rather conservative number of  the spin direction correlation measurements.

In analyzing the spin signatures of different growth channels, we have used rather simplified assumptions.
For example, we simply considered two extreme cases in analyzing MBH mergers:
wet mergers where gas disks are heavy enough to capture the MBH binaries onto them
and therefore enable a tight (remnant) MBH-host galaxy spin direction correlation;
dry mergers where there is no gas disk at all and MBH binary orbits are randomly
oriented with respect to the galaxy disks. Even a MBH merger is completely dry (in a gas free environment),
there could be a mild (remnant) MBH-host galaxy spin direction correlation,
because both spin directions are affected by the orbit direction of the pre-merger galaxy pair.
To accurately compute this correlation, one need to keep track of the evolution of MBH pairs driven
by all the processes outlined in Sec.~\ref{sec:merger} during multi galaxy mergers.
Exacting the meaningful initial conditions for the final inspiral and merger of the inner MBH binary from cosmological N-body simulations is particularly challenging.
To our knowledge, no self-consistent study of this problem is available.
In a more economical approach, coarse-grain simulation results of galaxy mergers are used as initial conditions of MBH binaries,
from which MBH binaries migrate inwards under different driven processes described with simplified prescriptions
which however lose track of both the MBH orbit direction in each merger
and the remnant BH spin direction in successive mergers \citep{Sayeb2020}. The growth channel Dry Mergers is distinguished
from other channels mainly by their different imprints on MBH spin magnitudes (see the 2nd row of Table~\ref{tab}),
therefore ignoring the  MBH-host galaxy spin direction correlation arising from Dry Mergers
does not undermine our conclusion.

In analyzing accretion,
we have assumed a Gaussian distribution $\mathcal N (1, 0.05)$ of MBH spins driven by coherent accretion. However, the realistic distributions in each channel may be considerably different. The bias or theoretical uncertainty in assessing the model distribution will inevitably affect the results of model selection. It is difficult to nail down all the theoretical uncertainties in this study, as it depends on many details of accretion that are hard to model from first principle:
the disk thickness, the magnetic field strength, the configuration of
magnetic fields lines and the matter emission properties of the inner disk.
More theoretical efforts along this direction are needed, otherwise in the
coming epoch of LISA some of our understanding of MBH formation may be limited
by the accuracy of modeling given all the astrophysical processes involved,
instead of the data uncertainty.

In reality, more than one dominant channels may play important roles, so that the detected data may imply a mixed distribution from various channels. Then the question becomes how to determine the mixing ratios of various channels based on LISA observations and corresponding electromagnetic counterpart measurements. There have been some efforts towards more accurately modeling the MBH growth
taking account of mixed channels and detailed astrophysics
\citep[see e.g.,][]{Barausse2012,Sesana2014,Kulier2015,Zhang2020,Bha2020, Sayeb2020}.
We expect similar discussion can be applied taking the full spin magnitude and MBH spin -galaxy spin correlation into account.

In this paper, we have shown the huge potential of probing the MBH growth via the MBH-galaxy spin direction correlations, in addition to the spin magnitude distribution.
As shown by \cite{Kuns2019}, host galaxies of stellar-mass binary BH mergers can be identified from combined observations of
a deci-hertz GW detector and a ground based detector to redshift $z\approx 0.3$, i.e., $\sim 400$ pairs of binary BH-host galaxy
would be identified per year (assuming a constant merger rate $60$ Gpc$^{-3}$yr$^{-1}$). If there is an non-negligible correlation between the orbital angular momentum of field-borne binaries and the rotation direction of their host galaxies, it will be interesting to explore
what we can learn from these stellar-mass systems.

\section*{Acknowledgements}
Z.P. and H.Y. are supported by the
Natural Sciences and Engineering Research Council of
Canada and in part by Perimeter Institute for Theoretical
Physics. Research at Perimeter Institute is supported in part
by the Government of Canada through the Department of Innovation,
Science and Economic Development Canada and by the Province of
Ontario through the Ministry of Colleges and Universities.

\bibliography{ms}

\end{document}